
\input epsf

\ifx\epsffile\undefined\message{(FIGURES WILL BE IGNORED)}
\def\insertplot#1#2{}
\def\insertfig#1#2{}
\else\message{(FIGURES WILL BE INCLUDED)}
\def\insertplot#1#2{
\midinsert\centerline{{#1}}\vskip0.2truein\centerline{{\epsfxsize=4.0truein
\epsffile{#2}}}\vskip0.5truecm\endinsert}
\def\insertfig#1#2{
\midinsert\centerline{\epsffile{#2}}\centerline{{#1}}\endinsert}
\fi

\input harvmac
%
%
%
%
\ifx\answ\bigans
\else
\output={
  \almostshipout{\leftline{\vbox{\pagebody\makefootline}}}\advancepageno
}
\fi
%
%
%
\def\mayer{\vbox{\sl\centerline{Department of Physics 0319}%
\centerline{University of California, San Diego}
\centerline{9500 Gilman Drive}
\centerline{La Jolla, CA 92093-0319}}}
%
%

%
%
%
%
\def\abstract#1{\centerline{\bf Abstract}\nobreak\medskip\nobreak\par #1}
%
%
%
%
\edef\tfontsize{ scaled\magstep3}
 \tfontsize  \tfontsize
 \tfontsize \font\titlei=cmmi10 \tfontsize
\font\titleis=cmmi7 \tfontsize \font\titleiss=cmmi5 \tfontsize
\font\titlesy=cmsy10 \tfontsize \font\titlesys=cmsy7 \tfontsize
\font\titlesyss=cmsy5 \tfontsize  \tfontsize
\skewchar\titlei='177 \skewchar\titleis='177 \skewchar\titleiss='177
\skewchar\titlesy='60 \skewchar\titlesys='60 \skewchar\titlesyss='60
%
%
%
%
%
\def\inv{^{\raise.15ex\hbox{${\scriptscriptstyle -}$}\kern-.05em 1}}
\def\lbar{{\lower.35ex\hbox{$\mathchar'26$}\mkern-10mu\lambda}} 

%
%
%
%
\def\dsl{\,\raise.15ex\hbox{/}\mkern-13.5mu D} 
\def\delsl{\raise.15ex\hbox{/}\kern-.57em\partial}
\def\Ksl{\hbox{/\kern-.6000em\rm K}}
\def\Asl{\hbox{/\kern-.6500em \rm A}}
\def\Dsl{\hbox{/\kern-.6000em\rm D}} 
\def\Qsl{\hbox{/\kern-.6000em\rm Q}}
\def\gradsl{\hbox{/\kern-.6500em$\nabla$}}
%
%
\def\lspace{\ifx\answ\bigans{}\else\qquad\fi}
\def\lbspace{\ifx\answ\bigans{}\else\hskip-.2in\fi} 
%
%
\def\boxeqn#1{\vcenter{\vbox{\hrule\hbox{\vrule\kern3pt\vbox{\kern3pt
        \hbox{${\displaystyle #1}$}\kern3pt}\kern3pt\vrule}\hrule}}}
%
%
\def\mbox#1#2{\vcenter{\hrule \hbox{\vrule height#2in
\kern#1in \vrule} \hrule}}
%
%
%
%
\def\CA{{\cal A}}   
   
   \def\CL{{\cal L}}

%
%
%
%
%

%

\def\bar#1{\overline{#1}}
\def\vev#1{\left\langle #1 \right\rangle}

\def\darr#1{\raise1.5ex\hbox{$\leftrightarrow$}\mkern-16.5mu #1}

%
%
\def\frac#1#2{{\textstyle{#1\over #2}}} 
%
%
%
%

\def\Im{\mathop{\rm Im}}
\def\Re{\mathop{\rm Re}}

\def\GeV{{\rm GeV}}

%
%
%
%

%
%
\def\ltap{\ \raise.3ex\hbox{$<$\kern-.75em\lower1ex\hbox{$\sim$}}\ }
\def\gtap{\ \raise.3ex\hbox{$>$\kern-.75em\lower1ex\hbox{$\sim$}}\ }
\def\gl{\ \raise.5ex\hbox{$>$}\kern-.8em\lower.5ex\hbox{$<$}\ }
\def\roughly#1{\raise.3ex\hbox{$#1$\kern-.75em\lower1ex\hbox{$\sim$}}}
%
%

%

%

\def\pl#1#2#3{{Phys. Lett. } {#1}B (#2) #3}
\def\prl#1#2#3{{Phys. Rev. Lett. } {#1} (#2) #3}
\def\physrev#1#2#3{{Phys. Rev. } {#1} (#2) #3}

\relax

\def\pone{\varphi_1}
\def\ptwo{\varphi_2}
\def\Re{{\rm Re\,}}
\def\Im{{\rm Im\,}}
\def\vev#1{\langle\, #1 \,\rangle}
\def\({\left(}
\def\){\right)}
\def\[{\left[}
\def\]{\right]}
\def\D{{\rm D}}
\def\sa{\sin\alpha}
\def\ca{\cos\alpha}
\def\cw{\cos\theta_W}
\def\sw{\sin\theta_W}
\def\b{\xi_V}
\def\c{\xi_A}

\noblackbox
\vskip 1.in
\centerline{{\titlefont{Flavour Changing Neutral Currents, }}}
\medskip
\centerline{{\titlefont{Weak-Scale Scalars and Rare Top Decays }}}
\bigskip
\vskip .3in
\centerline{Michael Luke and Martin J. Savage\footnote{$^{\dagger}$}
{SSC Fellow}}
\bigskip
\mayer
\bigskip\bigskip
\centerline{{\bf Abstract}}
\bigskip

We examine the decays $t\rightarrow c\gamma$ and $c Z^0$ in
the Standard Model with an extra scalar doublet
and no discrete symmetry
preventing tree-level flavour changing neutral currents.
The Yukawa couplings of the new scalars
are assumed to be proportional to fermion masses, evading bounds
on FCNC's from the light quark sector.
These rare top decays may be visible
at the SSC.

\vfill
\hbox{\hbox{ UCSD/PTH 93-01\hskip 2in March 1993} }
\eject

Flavour changing neutral currents (FCNC's) involving
the light quarks are known experimentally to be strongly suppressed,
placing stringent bounds on the nature of new flavour physics.
The gauge interaction
of the fermions in the standard model has a global
$\left[U(3)\right]^5$
flavour symmetry; this is broken to $\left[U(1)\right]^4$ by the
Higgs Yukawa
couplings, and the flavour symmetries are broken most strongly for
the heavier quarks.  If the observed pattern of
masses and mixing angles reflects an underlying pattern of symmetry
breaking, it is natural to expect that new flavour physics will also
respect this pattern of symmetry breaking.
Hence, it would couple
most strongly to the top quark, for which the flavour symmetry is
most strongly broken, while coupling only very weakly to the light
fermions.  This may allow the scale of new flavour physics
to be as low as the weak scale without contradicting current
experimental limits on FCNC's for light quarks.

A simple example is the Standard Model with extra scalars.
These new scalars will typically have tree-level flavour-changing
Yukawa couplings.  However, as has been discussed
by several authors \ref\chengsher{T.~P.~Cheng and M.~Sher,
\physrev{D44}{1991}{1461}\semi \physrev{D35}{1987}{3484}.}--%
\nref\weeman{M.J. Savage, \pl{266}{1991}{135}.}%
\nref\hou{W.-S.~Hou, \pl{296}{1992}{179}.}%
\nref\chang{D. Chang, W.S. Hou and W.Y Keung,
CERN-TH.6795/93 (1993).}%
\ref\anta{A. Antaramian, L.J. Hall and A. Rasin,
\prl{69}{1992}{1871}.}, if the Yukawa couplings of the
extra scalars are proportional to the fermion masses, low
energy limits on FCNC's
may be evaded because the flavour changing couplings
to the light quarks are small. The discrete symmetries which are
normally invoked in two-Higgs doublet models to
forbid tree level FCNC's \ref\glasb{S. Glashow and S. Weinberg,
\physrev{D15}{1977}{1958}.} are therefore unnecessary.
It has been suggested that these FCNC's may be looked for
in $b$ decays, and in particular the leptonic decay $B_s\rightarrow
\mu^+\mu^-$ may be observable at hadron colliders \weeman\anta.

If this philosophy is correct, one would expect to see the largest
effects of these FCNC's in $t$ decays.  The resulting rare
$t$ decays may be searched for at the SSC, where
$\sim 10^8$ $\bar tt$ pairs are expected to be produced per year.
FCNC's of gauge bosons in both the standard model
\ref\fca{V. Ganapathi, T. Weiler, E. Laermann, I. Schmitt and
P.M. Zerwas, \physrev{27}{1983}{579}}\nref\fcb{G. Mann and T. Riemann,
Annalen der Physik, 7 (1983) 334}\nref\fcc{M. Clements, C. Footman, A.
Kronfeld,
S. Narasimhan and D. Photiadis, \physrev{D27}{1983}{570}}\nref\fcd{A. Axelrod,
Nuc. Phys. {\bf B209} (1982) 349}\nref\fce{E. Ma and A. Pramudita,
\physrev{D22}{1980}{214}}-\ref\fcf{M.J. Duncan, \physrev{D31}{1985}{1139}}
and in two-Higgs doublet models with discrete symmetries
\ref\fcg{M.J. Savage, CALT-68-1496, (1988)}\nref\fch{C. Busch, Nuc. Phys. {\bf
B319}
(1989) 15}-\ref\fci{J. Hewett, S. Nandi and T. Rizzo, \physrev{D39}{1989}{250}}
have been studied extensively in the past.
In the Standard Model and two-Higgs doublet models (with ratios
of Higgs vevs of order one) the branching ratios for $t\rightarrow c Z^0$
and $t\rightarrow c\gamma$ are less than $O(10^{-11})$
\ref\grzad{B. Grzadkowski, J. F. Gunion and P. Krawczyk, \pl{268}{1991}{106}}
\ref\soni{G. Eilam, J.L. Hewett and A. Soni, \physrev{D44}{1991}{1473}}
and so would be unobservable at the SSC.
In this paper we will calculate the rates for
$t\rightarrow c\gamma$ and $t\rightarrow cZ^0$ in the Standard
Model with an extra scalar doublet, but without
the discrete symmetries which prevent tree-level FCNC's
in two-Higgs doublet models.\footnote{$^\dagger$}
{The mode $t\rightarrow c\gamma$ has also been recently discussed
in \ref\diaz{J. L. Diaz-Cruz and G. Lopez Castro,``CP Violation
and FCNC with the Top Quark'', to appear in Phys. Lett. B.}.}
We will not discuss the decay $t\rightarrow cg$ as the multi-jet
background from the primary decay mode $t\rightarrow b W^+$
will make this mode unobservable.
We will assume that the
new scalars are heavier than the top; the case of $m_h<m_t$ has
been considered in \hou, as
has the tree-level decay $h\rightarrow \bar t c$ for $m_h>m_t$.
We note that the latter process ($h\rightarrow W+$ jets)
has no clear signature and will be very difficult to see at
the SSC; in this case
loop effects such as those described here may be the only
way to see these couplings at a hadron collider.

We consider a model with two scalar $SU(2)_W$ doublets, $\pone$ and
$\ptwo$:
\eqn\phis{\pone=\pmatrix{\pone^+\cr \pone^0},\qquad
\pmatrix{\ptwo^+\cr \ptwo^0}}
with Lagrangian
\eqn\lag{\CL_\varphi=\D^\mu\pone^\dagger\D_\mu\pone +
\D^\mu \ptwo^\dagger\D_\mu\ptwo -V\(\pone,\ptwo\)}
where $V(\pone,\ptwo)$ is the most general potential consistent with
the gauge symmetries.  Since there is no global symmetry which
distinguishes the two doublets we may work in
a basis where only $\pone$ has a VEV without loss of generality:
\eqn\vevs{\vev{\pone}=\pmatrix{0 \cr v/\sqrt{2}},\qquad\vev{\ptwo}=0}
where $v=246\,\GeV$.  Three of the components of $\pone$ become the
longitudinal components of the $W^\pm$ and $Z^0$, while the spectrum
contains the charged scalars $\ptwo^\pm$, the neutral scalars $h^0$
and $H^0$ and the pseudoscalar $A^0$, where
\eqn\higgses{\eqalign{&H^0=\sqrt{2}\[\(\Re \pone^0 - v\) \ca +
\Re \ptwo^0 \sa\],\cr
&h^0=\sqrt{2}\[-\(\Re \pone^0 - v\) \sa +
\Re\ptwo^0 \ca\],\cr
&A^0=\sqrt{2}\[-\Im \ptwo^0\].}}
The mixing angle $\alpha$ is determined by the potential and is a
free
parameter of the model, as are the masses $m_{\ptwo^\pm}$,
$m_h$, $m_H$ and $m_A$.  The
Yukawa couplings to fermions are
\eqn\yukawa{\eqalign{\CL_Y=&\lambda^U_{ij}\bar Q_i \tilde \pone U_j +
\lambda^D_{ij} \bar Q_i \pone D_j + \lambda^E_{ij} \bar L_i \pone
E_j\cr
&+\xi^U_{ij}\bar Q_i \tilde \ptwo U_j +
\xi^D_{ij} \bar Q_i \ptwo D_j +
\xi^E_{ij} \bar L_i \ptwo E_j}}
where $\tilde\varphi_{1,2}\equiv i\tau_2\varphi_{1,2}^*$.
With the usual manipulations we write $\lambda^U$, $\lambda^D$ and
$\lambda^E$ in terms of the mass matrices $\sqrt{2}\,M^U/v$,
$\sqrt{2}\,
M^D V^\dagger/v$ and $\sqrt{2}\,M^E/v$,
where $V$ is the Kobayashi-Maskawa (KM) matrix.
We have used all our freedom to redefine the fermion fields to
diagonalise $M_U$, $M_D V^\dagger$ and $M_E$,
so the matrices $\xi^{U,D,E}$ are general $3 \times 3$
matrices:
\eqn\xis{\eqalign{&\xi^U=\pmatrix{\xi_{uu}&\xi_{uc}&\xi_{ut}\cr
\xi_{cu}&\xi_{cc}&\xi_{ct}\cr \xi_{tu}&\xi_{tc}&\xi_{tt}}\quad
\xi^D=\pmatrix{\xi_{dd}&\xi_{ds}&\xi_{db}\cr
\xi_{sd}&\xi_{ss}&\xi_{sb}\cr \xi_{bd}&\xi_{bs}&\xi_{bb}}\cr
&\xi^E=\pmatrix{\xi_{ee}&\xi_{e\mu}&\xi_{e\tau}\cr
\xi_{\mu e}&\xi_{\mu\mu}&\xi_{\mu\tau}\cr \xi_{\tau e}&\xi_{\tau\mu}&
\xi_{\tau\tau}}.}}
{\it A priori}, the $\xi$'s are all free parameters, and are constrained
in the light quark sector from the lack of observation of FCNC's.
Two possible forms for the scalar couplings, motivated by the
observed structures of the KM and fermion mass matrices,
are the Cheng-Sher
(CS) ansatz \chengsher
\eqn\csans{\xi_{ij}\sim {\sqrt{m_i m_j}\over v}}
and the Antaramian, Hall and Rasin (AHR) ansatz
(for Higgs vevs $v_1=v_2=v$) \anta
\eqn\ahrans{\eqalign{
&\xi^U_{ij}\sim {m_{U_j}\over v}\({m_{U_i} m_{D_i}
\over m_{U_j} m_{D_j}}\)^{1/4}\cr
&\xi^D_{ij}\sim {m_{D_j}\over v}\({m_{U_i} m_{D_i}
\over m_{U_j} m_{D_j}}\)^{1/4}\cr
&\xi^E_{ij}\sim {\sqrt{m_{E_i}m_{E_j}}\over v}.}}
In this work we will not make a specific ansatz for the $\xi$'s but will simply
assume that they are roughly proportional to powers of quark masses,
allowing us to retain only $\xi_{tt}$ and the flavour changing couplings
$\xi_{tc}$ and $\xi_{ct}$.  In general the $\xi$'s will
also be complex, introducing additional CP violation into the theory;
this will not be important for our results.

The general form of the amplitude $\CA\(t\rightarrow c+V\)$ where
$V=\gamma$ or $Z^0$ is
\eqn\mata{\CA = {1\over 16 \pi^2} \bar u_c\left( A\,\gamma^\mu
+ B\,\gamma^\mu\gamma_5 + iC\,\sigma^{\mu\nu}{q_\nu \over m_t} +
iD\,\sigma^{\mu\nu}\gamma_5 {q_\nu\over m_t}\right)u_t \epsilon_\mu,}
where $\bar u_t$, $u_c$ and $\epsilon_\mu$ are the incoming and outgoing
spinors and the gauge boson polarisation vector respectively.
The coefficient functions $A, B, C$ and $D$ are computed from
the diagrams in \fig\loopsg{Feynman diagrams contributing to the decays
$t\rightarrow c\gamma$ and $t\rightarrow cZ^0$}.   There are contributions
from both neutral and charged scalars in the intermediate states; note that
the charged scalar couplings are not proportional to the KM matrix in this
model. The expressions
for the coefficient functions are lengthy and are given in
the Appendix.  For the photon, electromagnetic gauge invariance implies
that the vector and axial vector form factors $A$ and $B$ vanish.
In terms of the coefficient functions the decay widths are
\eqn\widgam{\eqalign{
\Gamma (t\rightarrow c\gamma) & = {1\over 8\pi}m_t \left(|C|^2+|D|^2\right)\cr
\Gamma (t\rightarrow cZ^0) & = {1\over 16\pi m_t} \left(1-{m_Z^2\over m_t^2}
\right)\left({m_t^2\over m_Z^2}-1\right)
\left[   (m_t^2+2m_Z^2)(|A|^2+|B|^2) \right.\cr
&\left.  - 6m_Z^2 Re( A^*C-B^*D) + m_Z^2({m_Z^2\over m_t^2}+2)(|C|^2+|D|^2)
\right]}}
from which we compute the branching fraction by normalising to the rate for
$t\rightarrow bW$
\eqn\widtbw{\Gamma (t\rightarrow bW) = {G_F\over 8\pi\sqrt{2}} |V_{bt}|^2 m_t^3
\left(1-{m_W^2\over m_t^2}\right)
\left(1+{m_W^2\over m_t^2}-2{m_W^4\over m_t^4}\right).}

To avoid an excess of free parameters, we have plotted the branching
ratios as a function of a common scalar mass, $m_{\ptwo^\pm}=
m_h=m_H=m_A\equiv M$ in \fig\gam{Branching fraction for
$t\rightarrow c\gamma$ and $t\rightarrow c Z^0\rightarrow
c\,e^+e^-+c\,\mu^+\mu^-$
divided by the product of coupling constants
$(\xi_{tt}\bar\xi_{tc})^2$ plotted as a function
of a common scalar mass.
The solid, dashed and dot-dashed curves correspond to a top quark
mass
of 110, 150, 180 GeV respectively. }.
We have also set the flavour changing
couplings of the charged scalars, neutral scalars and pseudoscalar
to be equal to a generic coupling
constant $\bar\xi_{tc}$, where
$\bar\xi_{tc}\sim\xi_{tc}\sim \xi_{ct}\sim(\xi_{tc}\pm\xi_{ct})/2$.
These simplifications do not qualitatively change our results.  We have
multiplied our branching fraction to $Z^0$'s by the
$6.7\%$ branching fraction for $Z^0\rightarrow e^+e^-+\mu^+\mu^-$
\ref\pdg{Particle Data Group, \physrev{D45}{1992}{S1}.}, as the hadronic decays
of the
$Z^0$ have an insurmountable background from the dominant two-body
decay
of the top $t\rightarrow bW\rightarrow b+{\rm hadrons}$.

For an ansatz of the form \csans\  we
expect $(\xi_{tt}\bar\xi_{tc})^2\sim m_t^3 m_c/v^4
\sim 10^{-3}$.  The more complicated anzatz \ahrans\ gives
a similar estimate.  Given the expected $\sim 10^8$ $\bar t t$ pairs
expected per year at the SSC, this corresponds, for light scalars with
masses of order the $t$ quark mass, to at best only a few events
per year in the photon channel.   For larger couplings, the
expected signal of course dramatically improves.

It is useful to examine our results in the context of other limits
on the scalar masses which have been discussed in the literature.
The most stringent limits come from ${\bar B^0}-B^0$ and
${\bar K^0}-K^0$ mixing.  For the AHR ansatz \ahrans, these
give lower limits $M\gtap 400$ GeV and $M\gtap 500$ GeV
respectively\anta.  In a recent work \chang, Chang, Hou and Keung
have calculated the
two-loop contributions to $\mu\rightarrow e \gamma$ using the
CS ansatz \csans.  From the current upper limits on this decay mode
they placed a lower bound  of $M\gtap 200$ GeV.
Other processes place lower limits
on $M$ which are less than the $t$ quark mass and hence do
not constrain our results.  It is important to note, as pointed out
in \weeman\ and \chang, that these constraints arise from
couplings to the down-type quark and lepton sectors only.  It is
conceivable that flavour changing couplings are large
only in the up-type quark sector, for which the constraints
on $M$ are much weaker.  The possibility of $M\sim m_t$
with large flavour
changing couplings in the up sector is not excluded.
In the absence of any
compelling theoretical reason to prefer a particular choice of
$\xi^U$'s we feel it is best to simply regard them as free parameters to be
measured or constrained at the SSC.

In conclusion, we have calculated the branching ratios for
$t\rightarrow c\gamma$ and $t\rightarrow c Z^0$ in
the Standard Model with an extra scalar doublet, but
without discrete symmetries preventing tree-level FCNC's.
For flavour changing
couplings suggested by simple ansatzes for the Yukawa couplings
these decay modes would not be observable at the SSC, given the
constraints on the scalar masses from other processes.
However, these constraints may be evaded if flavour changing
couplings are larger in the $u$ quark sector than for the $d$
quark and lepton sectors.  In this case, the flavour changing
couplings and scalar masses are virtually unconstrained.
Consequently, sizeable branching ratios for $t\rightarrow c\gamma$
and $t\rightarrow c Z^0$ are possible.  To determine whether or not
these decays are observable at the SSC requires a thorough
background study; we have not attempted this in this work.
\bigskip

We wish to thank Ann Nelson and David Kaplan for numerous
discussions.   This work was supported in part by DOE
grant DE-FG03-90ER40546 and by National Science Foundation
Grant PHY-8958081.
MJS acknowledges the support of a Superconducting Supercollider National
Fellowship from the Texas National Research Laboratory Commission
under grant FCFY9219.

\appendix{A}{Expressions for the Coefficient Functions}

For simplicity, we set the mixing angle $\alpha$ to zero; the flavour changing
couplings then only involve $\ptwo$.
We split each coefficient up into contributions from graphs with
neutral scalars, graphs with pseudoscalars, graphs with both neutral
scalars and pseudoscalars and finally graphs with charged scalars:
$A=A^h + A^A + A^M + A^C$.
The contribution from graphs with only neutral scalars is given by
\eqn\matb{\eqalign{
A^h &= \int_0^1 dx \( -x\log{\beta^h\over \mu^2}+
\log\gamma^h\)\(\b^h a_V-\c^h a_A\)\cr
&\qquad +\int dx\,dy\, \left[ \( 1+
\log{\eta^h\over\mu^2}-{xy k^2\over\eta^h}\)
\(\b^h a_V+\c^h a_A\)\right.\cr
&\qquad\left.+{m_t^2(x+y-2)\over\eta^h}
\(\b^h a_V-\c^h a_A\)\right]\cr
B^h &= \int_0^1 dx \( -x\log{\beta^h\over\mu^2} +
\log\gamma^h\)\(\b^h a_A-\c^h a_V\)\cr
&\qquad +\int dx\,dy\, \left[\( -1-
\log{\eta^h\over \mu^2}+{xy k^2\over\eta^h}\)
\(\b^h a_A+\c^h a_V\)\right.\cr
&\qquad\left. +{m_t^2(x+y-2)\over\eta^h}
\(\b^h a_A-\c^h a_V\)\right]\cr
C^h & = m_t^2 \int dx\,dy\, \ {x\over\eta^h}
\(\b^h a_V-\c^h a_A\)+{2y-xy-y^2\over\eta^h}
\(\b^h a_V+\c^h a_A\)\cr
D^h & = m_t^2 \int dx\,dy\,\ {x\over\eta^h}
\(-\b^h a_A+\c^h a_V\)+{2y-xy-y^2\over\eta^h}
\(\b^h a_A+\c^h a_V\)\cr
}}
where we have defined the functions
\eqn\funa{\eqalign{\beta^h & =(1-x)^2m_t^2+xm_h^2\cr
                              \gamma^h & = { (1-x)m_t^2+ xm_h^2\over
(1-x)^2m_t^2 + xm_h^2}\cr
                              \eta^h & = (1-x-y)m_h^2 +
(x+xy+y^2)m_t^2-xyk^2 + i\epsilon}}
and the couplings
\eqn\consta{a_V^\gamma  ={2e\over 3}, \ \ \ a_A^\gamma = 0,\ \ \
a_M^\gamma  = 0,}
for the photon,
\eqn\constb{ a_V^Z  ={g\over 4\cw}(1-{8\over 3}\sw^2) , \ \ \
a_A^Z = -{g\over 4\cw}, \ \ \ \
a_M^Z = {g\over 2\cw}}
for the $Z^0$.   The products of Yukawa couplings are
\eqn\yukh{\b^h=\xi_{tt}(\xi_{ct}+\xi_{tc}^*)/4,\quad
\c^h=\xi_{tt}(\xi_{ct}-\xi_{tc}^*)/4}
 Finally, $k^2$ is the mass of the gauge boson and $\mu$ is the renormalisation
scale.  The sum of the graphs is of course finite and the result independent
of $\mu$.

The contribution from graphs with only neutral pseudoscalars is given
by
\eqn\matc{\eqalign{
A^A &= \int_0^1 dx \( -x\log{\beta^A\over\mu^2} -
\log\gamma^A\)\(\b^A a_V-\c^A a_A\)\cr
&\qquad +\int dx\,dy\, \left[\( 1+
\log{\eta^A\over\mu^2}+{xy k^2\over \eta^A}\)
\(\b^A a_V-\c^A a_A\)\right.\cr
&\qquad\left. -{m_t^2(x+y)\over \eta^A}\(\b^A a_V
+\c^A a_A\)\right]\cr
B^A &= \int_0^1 dx \( -x\log{\beta^A\over\mu^2} -
\log\gamma^A\)\(\b^A a_A+\c^A a_V\)\cr
&\qquad +\int dx\,dy\, \left[\( -1-
\log{\eta^A\over\mu^2}-{xy k^2\over\eta^A}\)
\(\b^A a_A-\c^A a_V\)\right.\cr
&\qquad\left.-{m_t^2(x+y)\over\eta^A}
\(\b^A a_A+\c^A a_V\)\right]\cr
C^A & = m_t^2 \int dx\,dy\,{-x\over\eta^A}
\(\b^A a_V+\c^A a_A\) - {y(x+y)\over\eta^A}
\(\b^A a_V-\c^A a_A\)\cr
D^A & = m_t^2 \int dx\,dy\,{x\over\eta^A}
\(\b^A a_A + \c^A a_V\) - {y(x+y)\over\eta^A}
\(\b^A a_A - \c^A a_V\)
}}
where
\eqn\funa{\eqalign{\beta^A & =(1-x)^2m_t^2+xm_A^2\cr
                              \gamma^A & = { (1-x)m_t^2+ xm_A^2\over
(1-x)^2m_t^2 + xm_A^2}\cr
                              \eta^A  & = (1-x-y)m_A^2 +
(x+xy+y^2)m_t^2-xyk^2 + i\epsilon} }
and
\eqn\yuka{\b^A=\xi_{tt}(\xi_{ct}-\xi_{tc}^*)/4,\quad
\c^A=\xi_{tt}(\xi_{ct}+\xi_{tc}^*)/4 .}

The graphs involving both the scalar and pseudoscalar particles
give:
\eqn\matd{\eqalign{
A^M &= a_M\int dx\,dy\, \(\c^{M1}\(-\log{\eta^{M1}\over\mu^2}
+{m_t^2(1-y-xy-x^2)\over\eta^{M1}}
\)\right.\cr
&\qquad\qquad\qquad\left.+\c^{M2}\(\log{\eta^{M2}\over\mu^2}+
{m_t^2(1-y+xy-2x+x^2)\over\eta^{M2}}\)\)\cr
B^M &= a_M\int dx\,dy\,\(\b^{M1}\(-\log{\eta^{M1}\over\mu^2}
+{m_t^2(1-y-xy-x^2)\over\eta^{M1}}
\)\right.\cr
&\qquad\qquad\qquad\left.-\b^{M2}\left(\log{\eta^{M2}\over\mu^2} +
{m_t^2(1-y+xy-2x+x^2)\over\eta^{M2}}\)\)\cr
C^M &=  a_M m_t^2 \int dx\,dy\,\c^{M1}{1-y-xy-x^2\over\eta^{M1}}
+\c^{M2}{1-y+xy-2x+x^2\over\eta^{M2}}\cr
D^M &= -a_M m_t^2 \int dx\,dy\,\b^{M1}{1-y-xy-x^2\over\eta^{M1}}
-\b^{M2}{1-y+xy-2x+x^2\over\eta^{M2}}
}}
where
\eqn\mate{\eqalign{ \eta^{M1} & =
(1-2x-y+x^2+xy)m_t^2+xm_h^2+ym_A^2-xyk^2+ i\epsilon\cr
 \eta^{M2} & = (1-2x-y+x^2+xy)m_t^2+xm_A^2+ym_h^2-xyk^2 + i\epsilon}}
and
\eqn\yukm{\eqalign{&\b^{M1}=\xi_{tt}(\xi_{ct}-\xi_{tc}^*)/4,\quad
\c^{M1}=\xi_{tt}(\xi_{ct}+\xi_{tc}^*)/4\cr
&\b^{M2}=\xi_{tt}(\xi_{ct}+\xi_{tc}^*)/4,\quad
\c^{M2}=\xi_{tt}(\xi_{ct}-\xi_{tc}^*)/4.}}

Finally, the contribution from graphs involving the charged scalar is
\eqn\matf{\eqalign{ &A^C =B^C= \xi^C \left[  -4a_R\int_0^1 dx\,x\log\beta^C
\right.\cr &\left.\qquad\qquad+4b_L\left( {1\over 2} + \int dx\,dy\,\left[
\log{\eta^{C1}\over\mu^2} -
{xyk^2\over\eta^{C1}}\right]\right)\right.\cr
 &\qquad\qquad\left.+2j_c  \int dx\ dy \left(  \log{\eta^{C2}\over\mu^2}
- {m_t^2y(1-x-y)\over\eta^{C2}}\right)
\right]\cr
&C^C = - D^C =  \xi^C m_t^2 \int dx\,dy\,y(1-x-y)\left( {4 b_L\over \eta^{C1}}
- {2j_c\over \eta^{C2}}\right)}}
where ,
\eqn\matg{\eqalign{&\xi^C=\xi_{tt}\xi_{tc}^*/4\cr
&\beta^C  = x(m_{\ptwo^\pm}^2-(1-x)m_t^2)\cr
&\eta^{C1}  = (1-x-y)(m_{\ptwo^\pm}^2 - xm_t^2) - xyk^2+i\epsilon\cr
&\eta^{C2}  = (x+y)m_{\ptwo^\pm}^2 -y(1-x-y)m_t^2 - xyk^2+i\epsilon,} }
and the couplings are
\eqn\math{a_R^\gamma  = {e\over 3},\ \ \ b_L^\gamma = -{e\over 6}, \ \ \
j_c^\gamma =
e}
for the photon and
\eqn\mati{a_R^Z  = -{gs_W^2\over 3c_W},\ \ \ b_L^Z = -{g\over
4c_W}(1-{2\over 3}s_W^2),  \ \ \
j_c^Z  = {g\over 2c_W}(1-2s_W^2) }
for the $Z^0$.

\listrefs
\listfigs

\insertplot{Figure 1}{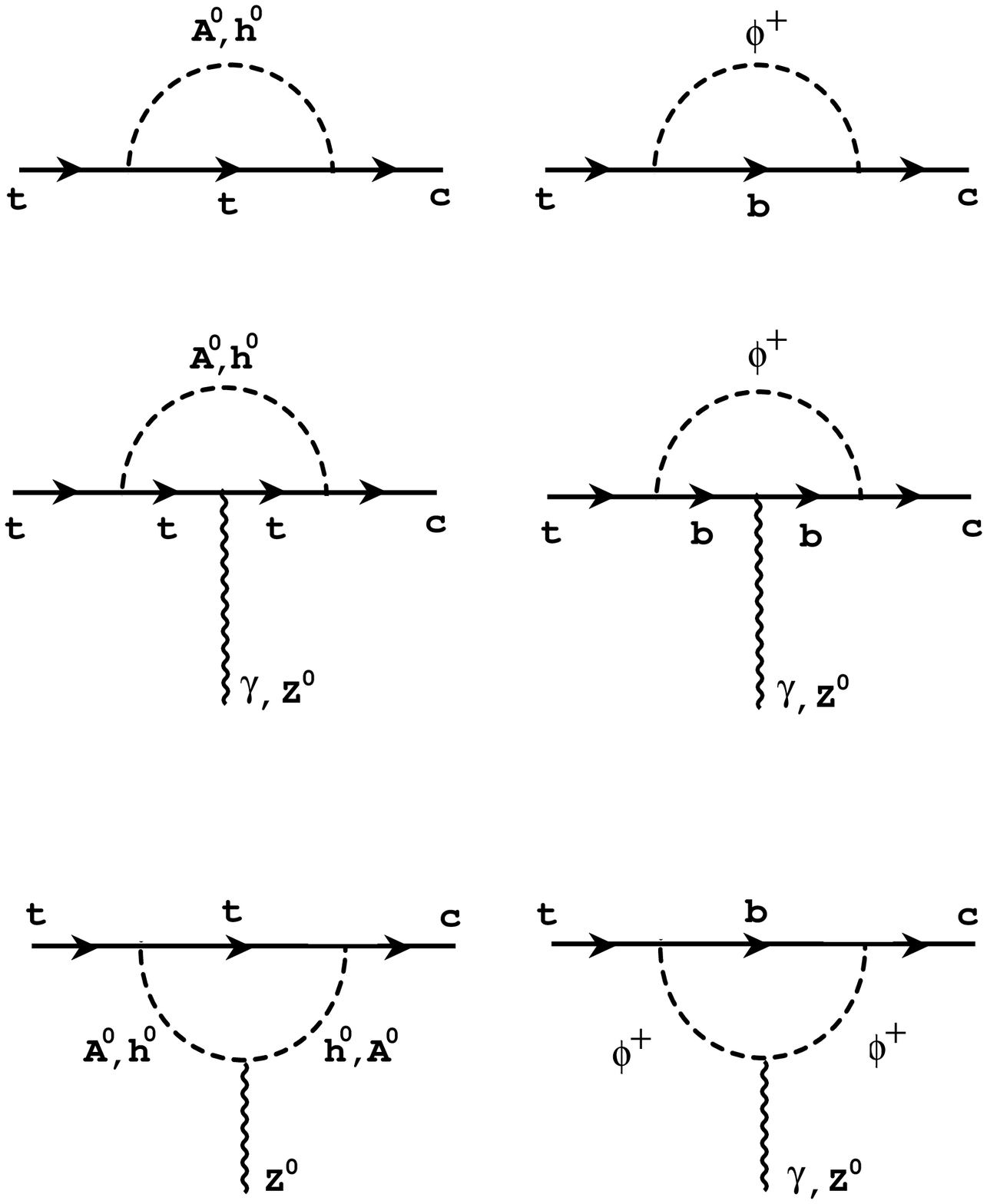}
\insertplot{Figure 2}{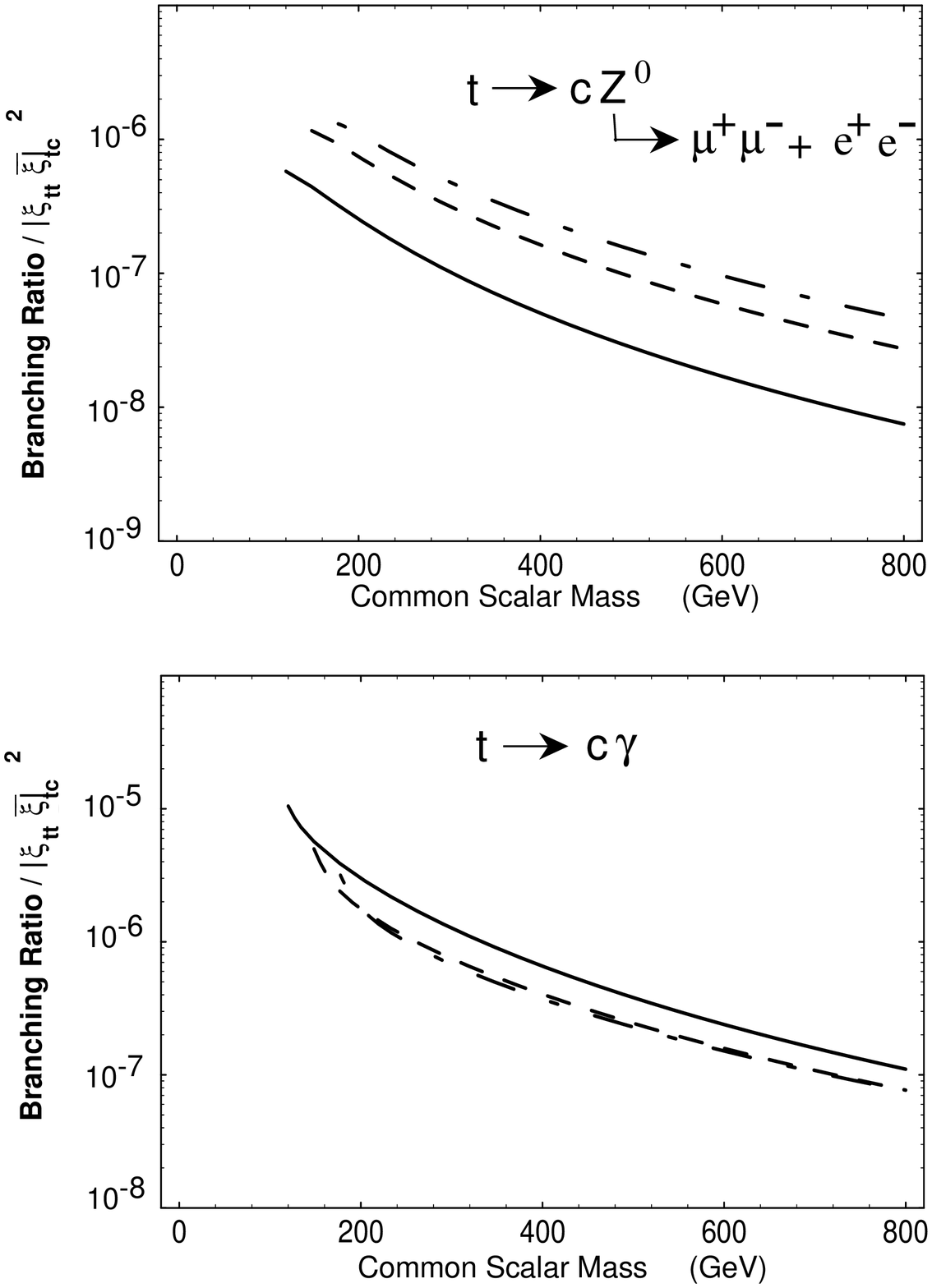}
\bye